\documentclass[
 aip,
 amsmath,
 amssymb,
 reprint,
]{revtex4-1}

\usepackage[english]{babel}
\usepackage[utf8]{inputenc}
\usepackage{float}
\usepackage[caption=false]{subfig}
\usepackage{graphicx} 
\usepackage{dcolumn} 
\usepackage{bm} 
\usepackage[utf8]{inputenc}
\usepackage[T1]{fontenc}
\usepackage{mathptmx}
\usepackage{etoolbox}
\usepackage{setspace}

\usepackage[letterpaper,top=2cm,bottom=2cm,left=3cm,right=3cm,marginparwidth=1.75cm]{geometry}

\usepackage{amsmath}
\usepackage{verbatim}
\usepackage{hyperref}
\hypersetup{colorlinks=true, allcolors=blue}

\def\Cplusplus{C\raisebox{0.5ex}{\tiny\textbf{++}}}
\newcommand{\abs}[1]{\lvert #1 \rvert}

\makeatletter
\def\@email#1#2{
 \endgroup
 \patchcmd{\titleblock@produce}
  {\frontmatter@RRAPformat}
  {\frontmatter@RRAPformat{\produce@RRAP{*#1\href{mailto:#2}{#2}}}\frontmatter@RRAPformat}
  {}{}
}
\makeatother
\begin{document}

\preprint{AIP/123-QED}

\title[Haptic Sensation-Based Scanning Probe Microscopy]{Haptic Sensation-Based Scanning Probe Microscopy: Exploring Perceived Forces for Optimal Intuition-Driven Control}
\author{M. Freeman}
\author{R. Applestone}
\author{W. Behn}
\author{V. Brar}
\email{vbrar@wisc.edu}

\affiliation{Department of Physics, University of
Wisconsin-Madison, Madison, Wisconsin 53706, United
States}

\date{\today}

\begin{abstract}
We demonstrate a cryogenic scanning probe microscope (SPM) that has been modified to be controlled with a haptic device, such that the operator can `feel' the surface of a sample under investigation.  This system allows for direct tactile sensation of the atoms in and on top of a crystal, and allows the operator to perceive, by using different SPM modalities, sensations that are representative of the relevant atomic forces and tunneling processes controlling the SPM.  In particular, we operate the microscope in modes of (1) conventional STM feedback, (2) energy-dependent electron density imaging, (3) q-plus AFM frequency shift based force sensing, and (4) atomic manipulation/sliding. We also use software to modify the haptic feedback sensation to mimic different interatomic forces, including covalent bonding, Coulomb repulsion, Van der Waals repulsion and a full Lennard-Jones potential.  This manner of SPM control creates new opportunities for human-based intuition scanning, and it also acts as a novel educational tool to aid in understanding materials at an atomic level.
\end{abstract}

\maketitle

\section{Introduction}

Scanning probe microscopy (SPM), including scanning tunneling microscopy (STM) and atomic force microscopy (AFM), is a powerful imaging technique that can be used to visualize the atomic structure of materials and the wavefunctions of electrons, as well as to manipulate individual atoms and molecules on the surfaces of materials.\cite{bian2021scanning, binnig1987scanning, chen2021introduction} These imaging methods work by using piezoelectric motors with sub-$\text{\AA}$ resolution to manipulate the position of an atomically sharp tip across a smooth material surface, where tip-surface contact is maintained using a feedback loop based on tunneling current, oscillation frequency shifting, or other measurable parameters that change with tip-sample separation.  Under standard operating conditions, the tip is raster scanned slowly across a surface and a full image is developed over the course of minutes to hours, depending on the specific imaging modality. More rapid, but less visual, measurements can be performed by continually scanning across the same path on the sample as some parameter is varied, or the tip can be simply parked over a specific spot of a surface to monitor changes as they happen in real time. During manipulation procedures, however, the tip is carefully lowered into the surface to either change its surface structure, or to pick up atoms/molecules or trap them beneath it.\cite{stroscio1991atomic, foster1988molecular}  Subsequent movement of atoms/molecules is performed by dragging them along the surface while moving the tip along a fixed path.\cite{lyo1991field, eigler1991atomic, ternes2008force}

In all of the above operational modes, live information about the tip height and feedback parameters can be monitored by the operator by inputting their values into an oscilloscope.  This allows for instantaneous changes in the tip or sample to be observed, and it also permits for precise placement of the tip relative to some surface feature by `blindly' moving the tip in sub-$\text{\AA}$ steps and monitoring --- for example --- the tip height as it is altered by individual surface atoms.  This later capability allows for manipulation to be performed without first obtaining sub-$\text{\AA}$ resolved images, which would be prohibitively time consuming.  During manipulation itself, a live feed of the scanning parameters is necessary for assessing if an atom/molecule is on the tip --- which appears as a change in height --- or if an atom/molecule is following the tip --- which can be known by observing the current noise;  while an oscilloscope can be used for such observations, it is more common to convert the tunneling parameters into an audible sound with, for example, the tip height and tunneling current represented by a particular pitch.\cite{weiss2008sounds, stroscio2004controlling}  This approach is often more intuitive and, in some cases, can provide a clearer picture of the tip and sample conditions.

Recently, commercial haptic feedback technologies have proliferated as a effective way in which to interact with a virtual space. Haptic feedback arms allow a user to move a cursor in a simulated 3D space, where contact with a virtual object results in a force imparted on the hand of the operator.  These devices are used extensively as tools for designing 3D virtual objects, performing robot-assisted surgeries, training for medical and dental operations, controlling robot arms in dangerous environments, and other applications.\cite{sheridan1992telerobotics}  They can be interfaced with any tool that exhibits motor-controlled $xyz$ motion and provides some positional feedback.  

Shortly after the innovation of piezo-controlled SPMs, efforts were made by several groups to interface them with haptic control arms as a way to `feel' the surfaces being probed.\cite{hatamura1990direct,hollis1990toward,taylor1997pearls,falvo1998nanomanipulation,sitti1998tele}  Those experiments, which included both STMs and AFMs, allowed users to feel coarse features on the surfaces of materials, including protruding quantum dots and step edges, with characteristic heights as small as 2nm.  Subsequent demonstrations showed that haptic-controlled AFMs could be used to sense and manipulate carbon nanotubes and large biomolecules, such as viruses, fibrin, and DNA.\cite{guthold1999investigation, reiner1999conceptual, guthold_controlled_2000, rubio2003force, sitti_teleoperated_2003, jobin2005versatile, falvo1998nanomanipulation} These experiments allowed the sensing of both lateral and vertical tip-sample forces as small as ~1nN. \cite{guthold_controlled_2000}  Later implementations also showed how haptic-controlled AFMs could be used to precisely position micron-sized beads,\cite{lim2009stable, xie_three-dimensional_2009, hou2012afm, schmid_human-operated_2012} sense semiconductor devices,\cite{lim2009stable} and sense the distance dependence of AFM tip-surface interactions. These later experiments were particularly useful for instructional purposes as the correlation between tip-surface and pen-hand interactions made it easier for students to understand forces between material interfaces. \cite{marchi2005interactive, tse2010haptics, bolopion_haptic_2013, millet2013haptics}

Most early haptic-SPM experiments were performed in ambient conditions, where it is difficult to maintain consistent tip and sample conditions, and where surface adhesion forces introduced from interfacial molecules can be significant.   Haptic experiments using optical tweezers provided a means of bypassing those challenges, with demonstrations showing reliable manipulation of trapped beads near and around barriers, and force sensing down to 1pN.\cite{pacoret_touching_2009,pacoret_haptic_2010,pacoret_invited_2013}  More recently, ultra-high vacuum (UHV) STM measurements --- where surface impurities were eliminated --- that included haptic arm control were performed on atomically smooth Au(111) surfaces with adsorbed C$_{60}$ molecules.\cite{perdigao2011haptic}  That work demonstrated that when STM feedback is controlled with tunneling current rather than tip-sample forces, the tunneling current could reliably be used to set the haptic arm force, allowing the operator to feel atomic step edges, and to even manipulate C$_{60}$ molecules. Other UHV experiments used camera tracking of hand movement to control an STM/AFM tip to manipulate PTCDA molecules on an Ag(111) surface.  \cite{green2014patterning, leinen2016hand}  Those experiments lacked force-feedback, but they demonstrated the effectiveness of hand-control in choosing optimal SPM tip trajectories.

Here we expand on those previous works by integrating a haptic feedback arm with a cryogenic, UHV SPM equipped with a quartz tuning fork sensor that can be used for both operation as a frequency modulated AFM and STM.  The ultra low temperature environment of the surface enables highly stable scanning conditions that we utilize to demonstrate several operational modes, including the linking of the haptic force to (1) tunneling current,  (2) changes in tuning fork frequency and (3) local electron density. We also explore the use of different feedback-force (`feeling') functions, which create different tactile sensations that can be used to mimic different atomic forces, and to optimize the ease of operation. Additionally, we demonstrate the ability to manipulate the positions of individual CO molecules with atomic resolution, such that the operator can feel as the molecules skip between atomic lattice sites.

\section{Methods}\label{sec:methods}

\begin{figure*}
\centering
\resizebox{\textwidth}{!}{\includegraphics{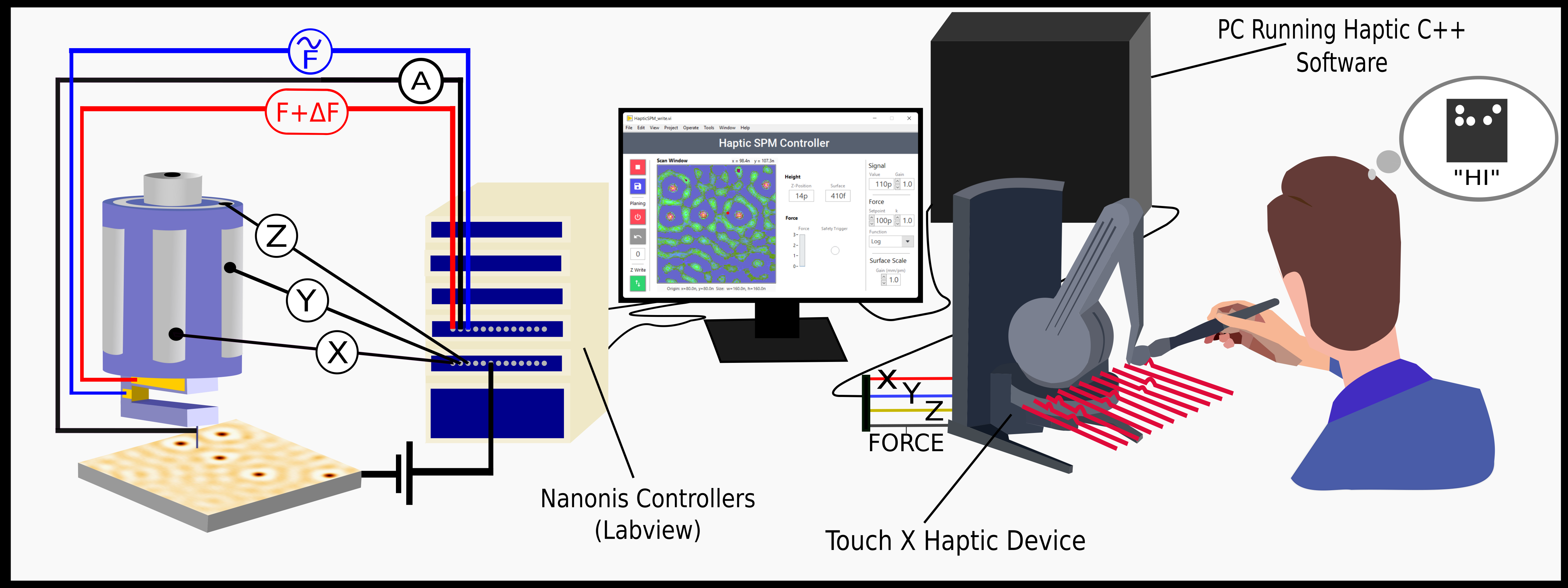}}
\caption{The operational setup used for the haptic-SPM controller. The haptic device, controlled via a dynamic-link library written in C/\Cplusplus{}, and the Nanonis control hardware, controlled via Nanonis's SPM control software, both connect to a desktop computer. Both of these interface via a LabVIEW program on the computer, which handles signal processing and the main GUI. The user moves the pen, and the xyz-positional data from the haptic device is mapped to control the xyz-position of the pen. Meanwhile, signals like the tip z-position, tunneling current, or cantilever frequency shift are converted into a force, applied to the haptic device, and felt by the user.  }
\label{fig:setup}
\end{figure*}

\subsection{Hardware and Software}
All measurements were performed in a commercial SPM (CreaTec GmbH) with Nanonis (SPECS GmbH) electronics and software.  The SPM is operated in UHV conditions at 4.5K using a qPlus quartz tuning fork sensor with a conductive PtIr wire tip that allows for dual AFM/STM imaging.   A Touch X haptic device (also called a haptic pen) from 3D Systems was used for all experiments. The device is operated via device drivers, calibration software, and the OpenHaptics 3.5 application programming interface (API) for the C/\Cplusplus{} languages. It possesses a workspace of 160 W $\times$ 120 H $\times$ 120 D mm, a maximum force output of 7.9 N, and a positional resolution of 1100 dots per inch (dpi) --- nearly double the resolution of haptic devices used in previous experiments.\cite{perdigao2011haptic} Such high resolution is ideal for obtaining smoother-feeling surfaces and forces.

There are three interconnected software components that allow the haptic device to communicate with the STM/AFM: (1) The Nanonis SPM control system, which is controlled through a LabVIEW-accessible API, and allows signal values (e.g., tip positional data, current, and oscillation frequency shift) and control parameters (e.g., tip position, tip bias) to be both read and controlled in LabVIEW; (2) a custom dynamic-link library developed using the OpenHaptics API, and coded in \Cplusplus{}, which allows positional and force parameters to be read and written to the haptic device via LabVIEW; and (3) a LabVIEW GUI program used to interface the two, processing signals from each device.  The LabVIEW GUI and dynamic link library can be found at \url{www.github.com/HapticSPM}.

The synthesis of these hardware and software components results in the complete haptic controller, as shown in Fig. \ref{fig:setup}. 

\subsection{Operational Basics}
\label{sec:oper}
There are two general modes of operation with which experiments were performed: \textit{read mode} and \textit{write mode}. In read mode, the haptic pen only controls the tip motion in the horizontal $xy$ plane, while the tip height is controlled via conventional STM/AFM feedback. The force on the haptic pen is then determined by reading a signal of interest (e.g., tip z-position, LDOS, cantilever frequency shift, etc.) while the feedback is on and referencing it to the z-position of the haptic pen; if the pen travels below a virtual, scaled version of the sample defined by the signal, it is pushed upwards. For example, if the feedback-determined tip-height is chosen as the signal, then the relative value of the actual SPM tip-height and the virtual tip height is used to set the force on the pen. The ‘feel' of the sample (i.e., force as a function of distance) is determined by a function of the form
\begin{equation}
    \label{eq:read}
    f(z_{pen}) =  \left\{
        \begin{array}{ll}
            k \abs{z_{pen} - a S_{fb}}  & \quad z_{pen} \leq a S_{fb} \\
            0 & \quad z_{pen} > a S_{fb}
        \end{array}
    \right. ,
\end{equation}

\noindent where $S_{fb}$ is the signal of interest with feedback on, $a$ is the gain setting controlled in LabVIEW, and $z_{pen}$ is the z-position of the haptic pen in mm. When the haptic pen is held `above' $a S_{fb}$, it experiences no force, but as the position of the pen hits $a S_{fb}$ and then goes below it, it experiences a spring force with an adjustable stiffness $k$. It is important to note, however, that the force-distance relation described in Eq. \ref{eq:read} need not be linear; the overall `feeling' of the sample can be changed by altering the form of this function as described later in Sec. \ref{sec:feel}. This mode of operation is similar to the `augmented reality mode' used in previous haptic-AFM implementations.\cite{vogl2006augmented, guthold_controlled_2000} 

In write mode, the haptic pen controls the horizontal xy-position, as well as the vertical z-position of the tip.  The force on the haptic pen is determined by the function $f(S)$, which is entirely dependent on the signal of interest $S$ without feedback, with the tip height set by the position of the pen. As an example, if the tunneling current is chosen as the signal of interest, then as the pen (and thus, tip) moves downward and approaches the surface, the tunneling current increases; this current is converted to an upwards force, which the user interprets as a hard surface.

The force functions discussed above are varied depending on the signal of interest, and are described explicitly in the next sections.

In normal use, a square region of interest is chosen on the surface of the sample using the standard Nanonis SPM control software. This region's location and size are then used by the haptic device to convert the millimeter-scale horizontal movement of the pen to the angstrom-scale horizontal movement of the tip. The workspace of the pen is confined using hard-coded, vertical walls to a $16$ cm $\times$ $16$ cm horizontal area in the middle of the pen's range of motion. This effectively confines the user of the haptic device to the square region of interest on the sample.  The sample surface is often slanted relative to the axes of movement of the scanning probe, and thus a planing algorithm must be used to `flatten' the surface, such that the horizontal movement of the pen is mapped to tip movement along the plane of the surface. This algorithm is implemented by having the user choose 3 points and fitting a plane to the force-defining signal at each of those points.

In both modes, it is possible to configure the perceived height of features on the surface by changing the desired level of positional scaling, which allows the user to feel differently-sized surface features. Thus, larger surface features like step-edges and smaller features like individual molecules/atoms are able to be felt in a more user-friendly way, without sacrificing the ability to feel only one or the other.

Finally, a drag force was implemented to smoothen the user's movements in an attempt to eliminate all but deliberate pen motions. This force was determined by the function $\mathbf{F}_{drag} = -c \mathbf{v}$, where $\mathbf{v}$ is the velocity of the haptic device cursor and $c$ is an arbitrary constant.
As an additional safety measure, the maximum speed of the STM/AFM tip could be configured such that the tip is only allowed to travel at less than a given speed. Beyond limiting unwanted rapid motions of the STM/AFM and the haptic device, this feature was found to be essential for fine-grained operations like molecular manipulation. 

\section{Modes of operation}

We tested our haptic-SPM controller on an atomically flat Cu(111) surface cleaned by repeated cycles of sputtering with Ar ions (500 V, ~7 $\mu$A) for 10 minutes and annealing around 400 C for ten minutes. The crystal was allowed to cool after annealing cycles and before each sputtering cycle. After loading the sample into the SPM, carbon monoxide molecules were deposited on the sample \textit{in situ} via a UHV leak valve.   The sample was analyzed using a variety of different haptic-SPM configurations, described below.

\subsection{STM Topography-Based Force}
\label{sec:stmread}

\begin{figure}[ht]
    \centering
    \resizebox{0.5\textwidth}{!}{\includegraphics{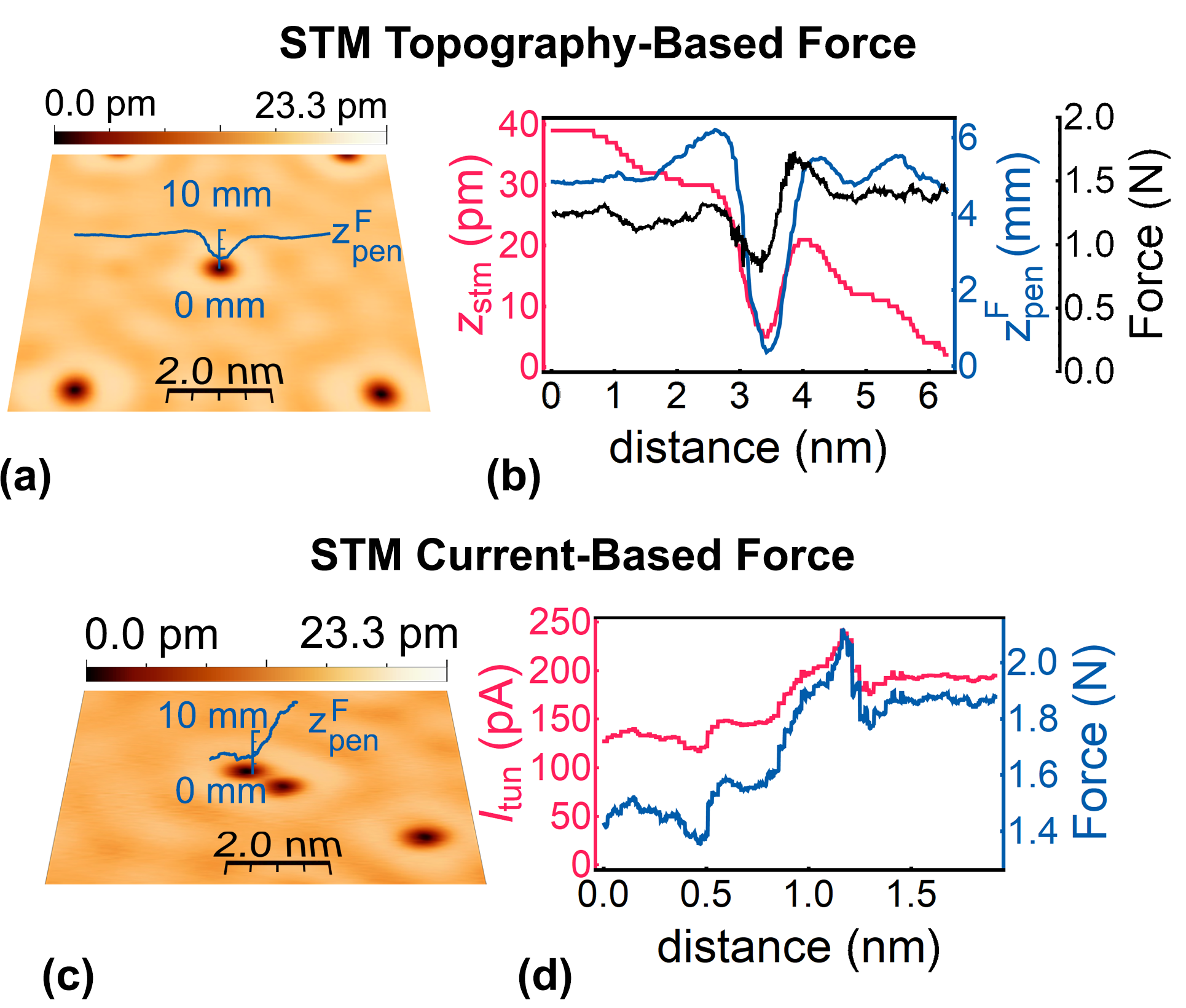}}
    \caption{An example of both `STM topography-based' (top) and `STM current-based' (bottom) operations. In (a) and (c) the pen position is plotted in three dimensions alongside the corresponding STM imagery. Linecuts that compare $z_{stm}$ and $z^F_{pen}$ as well as the force on the pen for topography-based operations are shown in (b).  Linecuts comparing the current vs. pen force as the tip traverses a molecule in current-based mode is shown in (d)}
    \label{fig:stm}
\end{figure}

In this operational mode, the STM feedback controller remains on, adjusting the z-position of the tip to maintain a constant `setpoint' current.  The haptic controller is operated in read mode, such that the haptic pen controls only the $xy$-position of the STM tip.  The force exerted on the pen is determined by Eq. \ref{eq:read} with $S_{fb}=z_{stm}$, where $z_{stm}$ is the height of the tip relative to the plane of best fit in nm. $a z_{stm}$ can be understood as the STM tip height read and projected into the virtual haptic space. 

Since the pen can move above $a z_{stm}$, its z-position may or may not correspond to the surface topography. Thus, $z_{pen}^F$ is used to distinguish the case in which the user is applying a constant finite downward force on the pen and the pen height would closely correspond to the surface topology.

Fig. \ref{fig:stm}a shows the three-dimensional path of the haptic device $z_{pen}^F(x_{pen},y_{pen})$ acquired using this mode of operation while transiting a carbon monoxide (CO) molecule on Cu(111).  The data is plotted alongside the STM surface topography, and Fig. \ref{fig:stm}b directly compares a linecut of the STM topography to $z_{pen}^F(x_{pen},y_{pen})$, showing good agreement. This indicates that the haptic pen is able to reliably feel sub-nanometer features.  

There are a noticeable discrepancies between the actual STM topography and the user-perceived surface, which originate from different sources.  First, the STM topographic data contains a slope that is representative of the actual slope of the sample, while the path of the haptic pen is flat.  This is due to the slope correction algorithm described in Section \ref{sec:oper}. Second, there is a 1 - 20 ms delay between when the STM signal value changes and when the haptic device writes the corresponding force. This delay is combined with a delay caused by the reaction time of the user, causing the haptic pen and SPM tip to be out-of-sync. Third, in order for $z^F_{pen}$ to match $z_{STM}$ the user must maintain a constant force on the pen during the entire movement, and small deviations are inevitable.  Finally, piezo drift differently affects the $z_{stm}$ and $z^F_{pen}$ profiles, since the former is obtained at a constant scan speed, and the latter's scan speed is controlled by the user, and generally non-constant. 

\subsection{STM Current-Based Force}
\label{sec:stmcurrent}

The next method of operation we describe implements a current-based force, where the STM feedback is turned off and the haptic controller is operated in write mode, meaning the pen controls the tip position in all dimensions.  The arm produces an upwards force proportional to the tip tunneling current, which increases with decreasing tip-sample separation. The force function used is 
\begin{equation}\label{eq:current}
f(I_{tun}) = I_{tun}/I_{set}
    \end{equation}
where $I_{tun}$ is the tunneling current, $I_{set}$ is an arbitrary current chosen by the user, which the pen tends to maintain during use.

We found setting the tip force directly to the tunneling current created difficult operating conditions when feeling large features, where small changes in tip height led to large changes in force that the haptic arm could not produce at sufficient rates to maintain stable tunneling;  it was also difficult to avoid tip crashes since the maximum force of the haptic pen could easily be reached with small changes in the tip height. To mitigate these problems, the vertical z-position of the pen was further processed using the equation
\begin{widetext}
\begin{equation}
\label{eq:pos_scal}
    z_{stm}(z_{pen}) = \left\{
            \begin{array}{ll}
                b (\exp{(\frac{1}{b}[z_{pen} - z_{thresh}])}-1)+z_{thresh}  & \quad z_{pen} \leq z_{thresh} \\
                z_{pen} & \quad z_{pen} > z_{thresh}
            \end{array}
        \right.
    \end{equation}
\end{widetext}

\noindent where $b$ is a scaling constant (nm), and $z_{thresh}$ ($nm$) is the z-position at which the tunneling current surpasses a predetermined current threshold (chosen to be about 20 pA in experiments shown here). Eq. \ref{eq:pos_scal} acts as a configurable `floor' past which the STM tip can only travel a few tens of picometers, regardless of how much further the haptic pen position $z_{pen}$ travels. This prevents problems feeling large features by dramatically decreasing the degree to which the haptic arm moves the pen towards the surface after a certain threshold height is met.  This correction also assists the operator in maintaining a constant tip height at $z_{thresh}$, since the transition to different functional forms on either side of $z_{thresh}$ can be easily felt by the user. When the composition of Equations \ref{eq:current} and \ref{eq:pos_scal} is considered, force as a function of pen height is at first exponential, then becomes linear as the force approaches the maximum force. 

A representative contour of $z_{pen}^F(x_{pen},y_{pen})$ obtained using a current-based force is shown in Fig. \ref{fig:stm}c, along with the corresponding topography obtained using standard STM imaging.  We find that the topography is less-well reproduced in comparison to using the topography-based force described in the previous section, and we attribute that difference to the 1 - 20ms communication delay between the haptic pen and STM, which makes it difficult for the user to maintain a perfectly constant current.   Fig. \ref{fig:stm}c, meanwhile, compares the force on the haptic pen against the tunneling current as the user traverses the tip across an individual CO molecule.  Here we find good agreement, with the current/force decreasing as the tip moves into the CO-defined dip (due to the tip moving further from the surface as it goes off an edge) and increasing as it moves out.

\bigskip
\bigskip

It is found that the topography-based mode described in Section \ref{sec:stmread} is easier to control and acquires better data than the current-based mode.  In both modes, the user attempts to maintain a constant downward force on the pen, establishing a feedback loop between the human, the haptic device, and the STM. However, in the current-based mode, human errors and inefficiencies in this loop affect the tip $z$-position and, by extension, the signal that sets the force, which can feel erratic.  Such processes include slow reaction times of the user, random hand movements, communication delay between the haptic pen and the STM, and microscopic changes to the SPM tip or sample.  On the other hand, in topography-based mode the tip $z$-position is controlled via electronic feedback, which is unaffected by user-error, and can quickly correct for changes in tunneling conditions (the typical feedback time constant is 30$\mu$s).  An additional caveat of the current-based mode is that it is necessary to know some approximate parameters of the signal of interest (i.e. how the signal varies with tip-sample separation) before operation in order to appropriately scale the position and force mapping --- this is not necessary when operating in topography-based mode.

Despite the above drawbacks, current-based force imaging is significantly more useful when a 4-dimensional understanding of the signal of interest is desired ($x$, $y$, $z$, and $I_{tun}(x,y,z,V)$); the topography-based mode is only able to provide information about a single, 3-dimensional contour of the signal of interest ($x$, $y$, and $z_{stm}$), and is thus more limited in its capacity to convey information about the surface.  For example, current-based imaging could allow the operator to feel changes in the electron decay constant associated with different bands\cite{stroscio1986electronic} or different molecular orbitals. \cite{stroscio1995tunneling, calleja2004contrast, picone2010atomic, tange2010electronic, woolcot2012scanning, takahashi2016orbital}

\begin{figure}[ht]
    \centering
    \resizebox{0.45\textwidth}{!}{
    \includegraphics{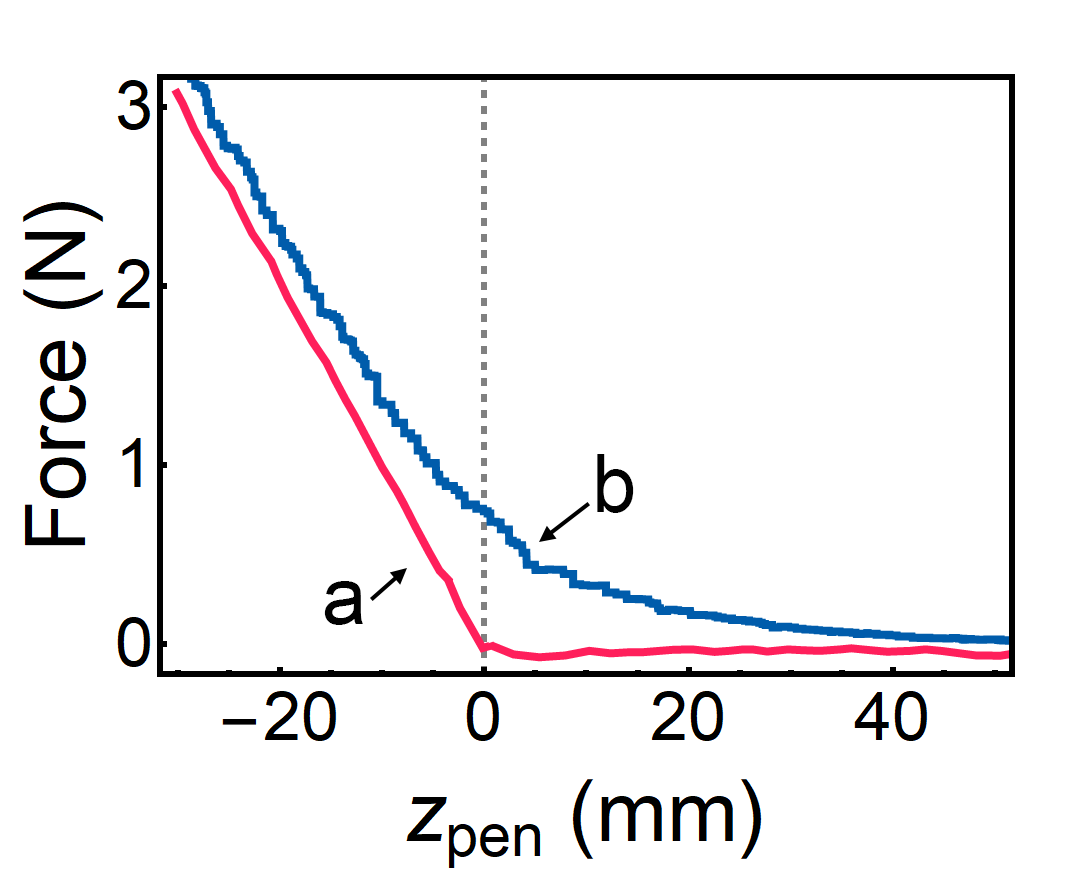}}
    \caption{Force as a function of $z_{pen}$ for the (a) STM topography-based force and the (b) STM current-based force.}
    \label{fig:plunge}
\end{figure}

Fig. \ref{fig:plunge} shows a comparison of force as a function of $z_{pen}$ when lowering the pen into the surface (taken to be at $z_{pen} = 0$ mm). The current-based force exhibits a relationship initially indicative of the tunneling current's exponential dependence on surface proximity and then becomes linearized at $z_{pen} = z_{thresh} = 0$ mm due to Eq. \ref{eq:pos_scal}, while the topography-based force exhibits a simple, linear relationship.

\subsection{STM LDOS-Based Force}
\label{sec:stmLDOS}

The tip-sample distance in an STM is typically maintained by using the tunneling current as the feedback parameter, however, the force on the haptic arm can be defined by any measurable value.  For example, the energy dependent local density of electronic states (LDOS) is one property that is commonly measured in STM experiments to provide direct visualization of electron wavefunctions and interference patterns.  The LDOS at any particular point of the sample is proportional to the tip-sample conductance at the tunneling voltage, and is commonly measured using a lock-in amplifier.  By converting the lock-in signal to a force written to the haptic controller operating in read mode, the user is able to feel the voltage-dependent local electron density as it varies across the sample.  

\begin{figure}[ht]
    \centering
    \resizebox{0.45\textwidth}{!}{
    \includegraphics{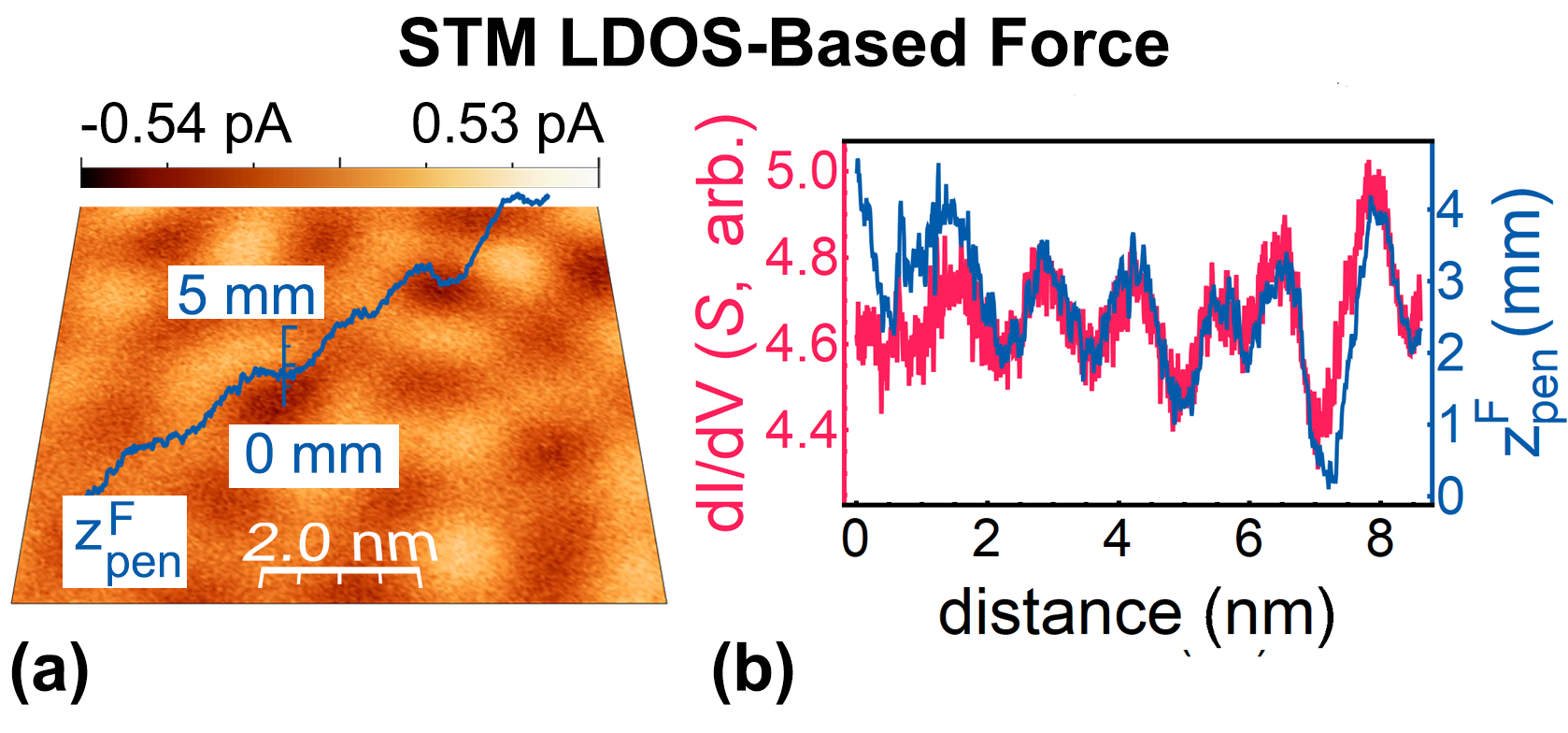}}
    \caption{An example of the LDOS-based mode in which the haptic pen was operated. The pen position is plotted in three dimensions alongside corresponding LDOS imagery is shown in (a); a direct linecut comparison is shown in (b).}
    \label{fig:ldos}
\end{figure}

This mode of operation uses Eq. \ref{eq:read} with $S_{fb} = LDOS_{stm}(V)$, and the gain variable $a$ having units of mm/V (the lock-in outputs a voltage that is proportional to the sample conductance).  Thus, the contour felt by the operator, $z_{pen}^F(x_{pen},y_{pen})$, is defined by the local sample conductance, which depends on the tunneling bias.  An example is shown in Fig.  \ref{fig:ldos}a, where $z_{pen}^F(x_{pen},y_{pen})$ of a particular path is plotted on top of an LDOS image of the Cu(111) surface, which reveals standing waves created by the interference pattern of the copper surface state.  Fig. \ref{fig:ldos}b, meanwhile, directly compares the trajectory of the haptic pen with the corresponding linecut of the LDOS image, showing excellent agreement.
 
We note that, unlike STM-derived topography, the LDOS changes dramatically with the applied bias, and those changes reveal the rich behavior of the electrons in the material.  In the case of Cu(111), LDOS images taken at higher/lower sample biases reveal shorter/longer wavelength standing wave patterns due to the different momenta of the participating electrons.\cite{crommie1993imaging}  Moreover, LDOS images of molecules can reveal the internal structure of the different molecular orbitals. \cite{soe2009direct, repp2005molecules, itatani2004tomographic, lu2003spatially}  These properties are all accessible to haptic force based sensing.

\begin{figure}
    \centering
    \resizebox{0.45\textwidth}{!}{\includegraphics{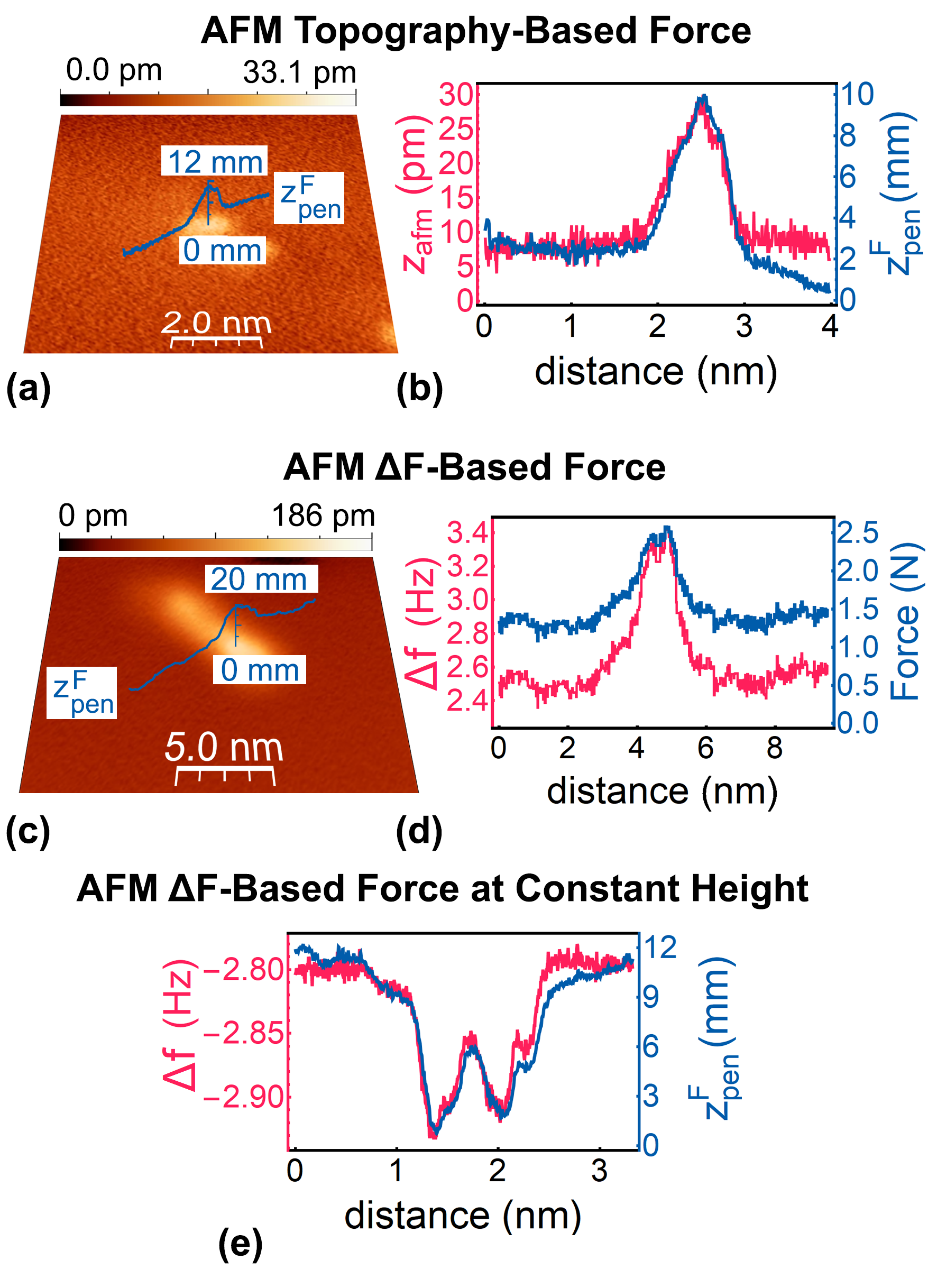}}
    \caption{An example of the three AFM-based modes in which the haptic pen was operated. The pen position  plotted in three dimensions alongside the corresponding AFM imagery is shown in (a) and (c); a direct linecut comparison is shown in (b) and (e), while (d) compares frequency shift and force.}
    \label{fig:afm}
\end{figure}

\subsection{AFM Topography-Based Force}\label{sec:afmheight}

In this mode of operation, the feedback of the system remains on, and the tip-sample distance is maintained at a constant height by linking the feedback to the frequency shift of the AFM cantilever ($\Delta f$), which changes depending on the tip-sample distance, and the specific tip-sample interactions (i.e. electrostatic, Van der Walls, etc...).    The haptic controller is operated in read mode, with $x$ and $y$ positions of the AFM tip controlled directly via the pen, and the force on the pen is set by Eq. \ref{eq:read}, where $S_{fb} = z_{afm}$. The cantilever oscillation phase was 2.47 degrees, and the oscillation excitation voltage was 2.97 mV.

Figure \ref{fig:afm}(a) shows the three-dimensional path of the haptic pen with constant force $z_{pen}^F(x_{pen},y_{pen})$ acquired using an AFM topography based force while transiting over a CO molecule on Cu(111).  The data is plotted atop the associated AFM image, and Fig. \ref{fig:afm}(b) directly compares $z_{pen}^F(x_{pen},y_{pen})$ to a linecut of the topography.  

This mode of operation performs similarly to the STM topography-based mode described in Sec. \ref{sec:stmread}, however, the perceived properties are fundamentally different since the STM is sensitive to conductivity and the AFM is sensitive to interatomic forces.  These differences are apparent in the measurements of CO molecules: they are felt as protrusions in AFM topography mode since they physically protrude from the surface and interact with the AFM tip, while they are perceived as dips in STM topography mode since the locally lower the surface conductivity.

\subsection{\texorpdfstring{AFM $\Delta f$-Based Force}{AFM Frequency Shift-Based Force}}
\label{sec:afmwrite}

In this configuration, the pen was operated in write mode, using a force based on the frequency-shift of the AFM cantilever. The function used was
\begin{equation}\label{eq:dF}
f(\Delta f) = \left\{
        \begin{array}{ll}
            0  & \quad \Delta f \leq \Delta f_{setpoint} \\
            k \log{(\Delta f/\Delta f_{setpoint})} & \quad \Delta f > \Delta f_{setpoint}
        \end{array}
    \right. ,
    \end{equation}
where $\Delta f$ is the measured frequency-shift of the cantilever, $\Delta f_{setpoint}$ is a value chosen by the user at which the pen begins to feel a force, and $k$ is a scalar which sets the gain.  We note that if the force on the pen is linked directly to $\Delta f$, the applied force on the pen is maxed out too quickly and the operator can easily crash the tip.  Using the logarithmic, step-like form shown in Eq. \ref{eq:dF} makes it easy to hold the tip at a particular setpoint, while still enabling tip-sample force perception as the tip is lowered further.

Force perception obtained while transiting a molecule in this mode is shown in Fig. \ref{fig:afm}c, while Fig. \ref{fig:afm}d directly compares the perceived force to $\Delta f$.  The major benefit of this mode of operation is that the pen force is directly linked to the actual physical force between the tip and sample (which establishes $\Delta f$).  This allows the user to feel the different forces that determine the AFM tip-sample interaction; this ability has previously been shown to be valuable in educational settings.\cite{marchi2005interactive,tse2010haptics,bolopion_haptic_2013,millet2013haptics}

It is found that the comparative usefulness between the AFM topography-based mode and the AFM $\Delta f$-based mode is identical to the corresponding STM modes, as discussed at the end of Section \ref{sec:stmcurrent}.  In particular, topography-based operation is more stable, while $\Delta f$-based scanning provides more opportunities for studying the surface properties via tactile sensation.

\subsection{\texorpdfstring{AFM $\Delta f$-Based Force at Constant Height}{AFM Frequency Shift-Based Force at Constant Height}}
\label{sec:afmfreq}

The last mode of operation we demonstrate is frequency-shift based, for which controller feedback was turned off, but the tip is fixed to move in a flat plane that is established by fitting topographic data obtained first by imaging the surface in standard feedback mode. Local changes in surface topology raise or lower the frequency shift from resonance of the AFM cantilever and this signal was converted to a force on the haptic device in read mode.  This method of imaging --- which is especially effective when $\Delta f$ is a non-monotonic function of $z_{afm}$ --- has been shown to be effective in obtaining atomically-resolved images of surfaces and, for CO functionalized tip, in imaging atomic bonds withing molecules. \cite{gross2009chemical, giessibl2019qplus, giessibl2003advances} 

The force on the haptic pen transiting two CO molecules in this mode is shown in Fig. \ref{fig:afm}e, plotted atop the associated direct measurement of $\Delta f$ from the Nanonis software.  We note that here we plot the raw value of $\Delta f$ rather than the absolute value, which would reveal peaks instead of dips.

\begin{table}[h]
\renewcommand{\arraystretch}{1.9}
\caption{\label{tab:table1}The specific functions used to configure how the surface feels. The constants $\rho$, $\tau$, $\sigma$, and $\epsilon$ were all chosen to provide the a force curve confined to a 30 mm range between zero force and the maximum force (3 N).  }
\begin{ruledtabular}
\begin{tabular}{lc}
Force&$\varphi(r)$\\
\hline
Linear & $\rho r$\\
Covalent & $\exp(\rho r-\tau)-\epsilon$\\
Coulomb & $\frac{\sigma}{(\rho r - \tau)^2}-\epsilon$ \\
Van der Waals & $\frac{\sigma}{(\rho r - \tau)^7}-\epsilon$\\
Lennard-Jones & $-4\epsilon \left( \frac{6\sigma^6}{(\rho r - \tau)^7}-\frac{12\sigma^{12}}{(\rho r - \tau)^{13}} \right)$\\
\end{tabular}
\end{ruledtabular}
\end{table}

\begin{figure}[h]
    \centering
    \resizebox{0.45\textwidth}{!}{
    \includegraphics{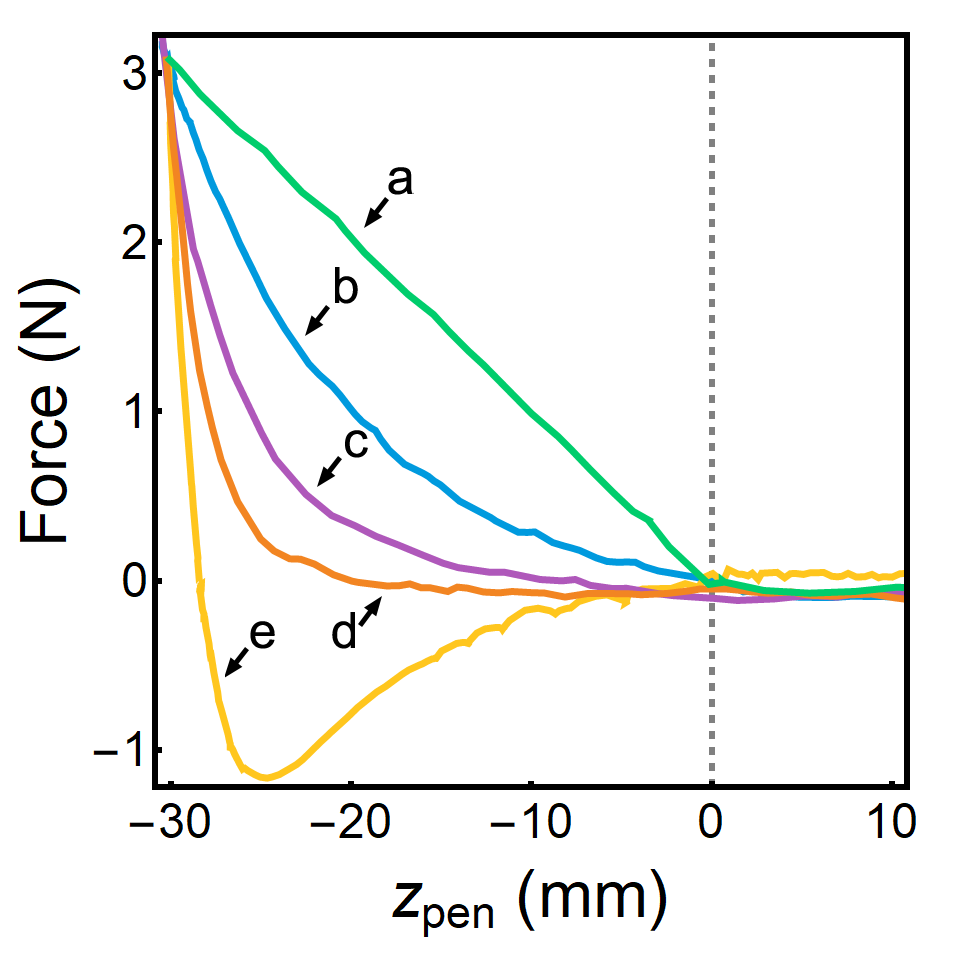}}
    \caption{Force as a function of $z_{pen}$ for each force investigated, including (a) linear force, (b) covalent bonding force, (c) Coulomb force, (d) Van der Waals force, and (e) Lennard-Jones force. All curves are shifted such that the surface is at $z_{pen} = 0$ mm. }
    \label{fig:ffplunge}
\end{figure}

\section{Feeling functions}
\label{sec:feel}

\begin{figure*}
\resizebox{\textwidth}{!}{\includegraphics{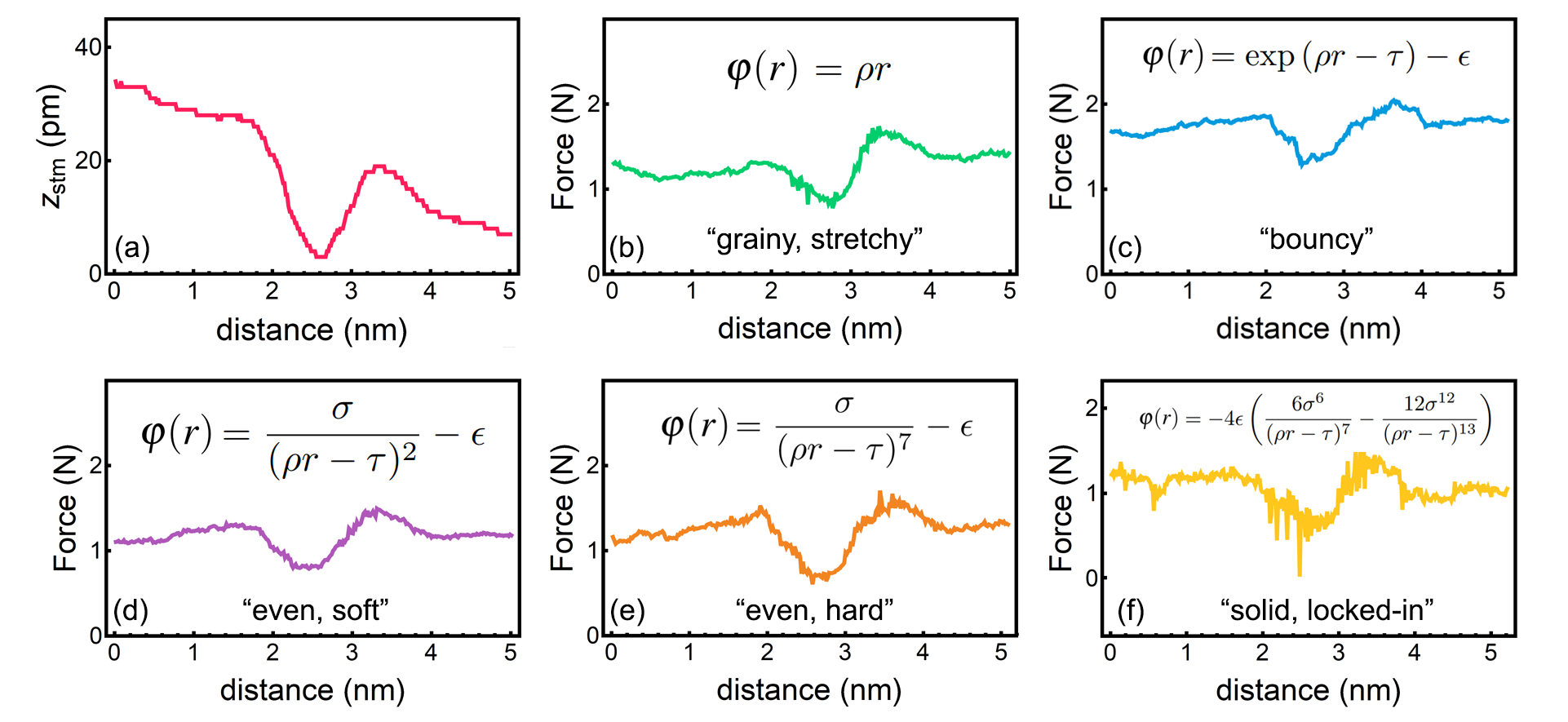}}
\caption{An example of the forces experienced while transiting a CO molecule using different 'feeling functions'. A linecut of the molecule imaged using conventional STM is shown in (a); a comparable path was taken with the haptic pen applying a (b) linear force, (c) covalent bonding force, (d) Coulomb force, (e) Van der Waals force, and (f) Lennard-Jones force. The subjective feel corresponding each function is written under each plot.}
\label{fig:forcefeelings}
\end{figure*}

Determining how the surface feels when using the haptic controller is an important parameter that is decided by the subjective preference of the operator and/or a distance-dependent SPM signal.  In absolute terms, the feel, or 'feeling function' ($f(z_{pen})$), is set by how the force exerted by the haptic pen depends on the pen height, but that dependence can be varied over a wide range of parameter space. In sections \ref{sec:stmread}, \ref{sec:stmLDOS}, and \ref{sec:afmheight}, $f(z_{pen})$ was set to be linear and spring-like, and independent of the actual SPM signal dependence on height;  Meanwhile, in Secs. \ref{sec:stmcurrent}, \ref{sec:afmwrite}, and \ref{sec:afmfreq} $f(z_{pen})$ was directly linked to the raw SPM signal, or a logarithmic/exponential function of it.  There is, however, substantial leeway in choosing a 'feeling function' that is engineered to impart sensations that deviate from the SPM signal vs. height function, with the only technical constraint being that it must prevent drastic force vs. tip height dependencies which can lead to unstable operation.  Otherwise, it can be chosen to either provide the best feel to the user, or to mimic a particular interatomic force.  In the later case, the force felt on the pen would represent the forces that a theoretical atom or molecule would feel if scanned across the surface.

\begin{figure*}
\resizebox{0.85\textwidth}{!}{\includegraphics{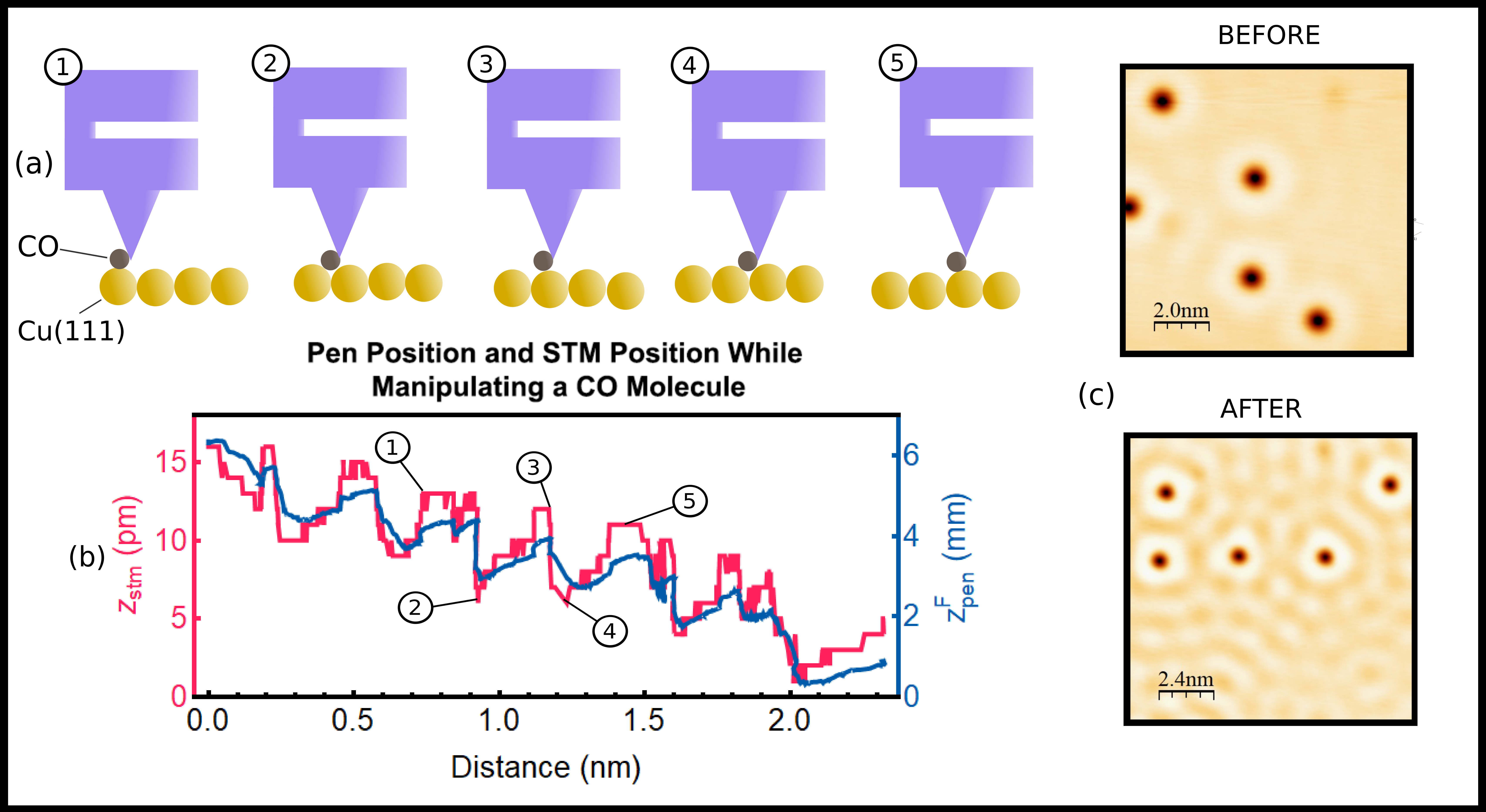}}
\caption{An example of controlled manipulation of CO on Cu(111) using the haptic device. Shown is (a) an illustration of a CO molecules being laterally dragged over the Cu(111) lattice by an STM tip (b) linecuts of $z^F_{pen}$ and $z_{stm}$ obtained during the lateral dragging process, which show atomic corrugations due the preferred binding of the CO molecule to the top of the Cu atoms, and (c) before and after images of the same area, in which CO molecules were moved to spell out `hi' in Unified English Braille. }
\label{fig:manip}
\end{figure*}

The idea of exploring different feeling functions is also motivated by studies that show that the psychometric function for compliance discrimination (i.e., perceived stiffness as a function of the actual stiffness) is roughly logarithmic, and on the order of N/mm.\cite{jones2012application} This seems to suggest that force-functions will yield significantly different perceptions than what might be initially predicted; just because force as a function of distance is exponential, for example, does not mean the surface will `feel exponential' to the user.

To investigate the effect of modifying $f(z_{pen})$ on the haptic pen operation, we used different `feeling functions' while probing the properties of an individual CO molecule.  All measurements were performed in the mode of STM Topography-Based Force (Sec. \ref{sec:stmread}), with Eq. \ref{eq:read} modified to mimic different interatomic forces.  The `feeling functions' used were of the form:

\begin{equation}
f(r) = \left\{
        \begin{array}{ll}
            \varphi(r) & \quad r \geq 0 \\
            0 & \quad r < 0
        \end{array}
    \right.
    ,
\end{equation}

\noindent where $r \equiv a z_{stm} - z_{pen}$ is defined to be the penetration distance, and $\varphi (r)$ takes on the various forms as described in Table \ref{tab:table1}.  These feeling functions impose sensations on the hand of the operator that  simulate the forces experienced by different species of molecules interacting with the surface, depending on the valence state, bond structure, charge, and size.  These different functions also modify the user's ability to control the haptic pen as they change the apparent shape and `hardness' of surface features, which affects the user's reaction time.  

In order understand the effect of these different feeling functions, we plot the the force experienced by the user as a function of pen height in Fig. \ref{fig:ffplunge} for each individual functional form.  In Figure \ref{fig:forcefeelings} we plot the perceived force using each function while transiting a CO molecule.  We note that while the magnitude of force fluctuations is similar between different feeling functions, the tactile sensation are different, as described in the figure.  For example, using a full Lennard-Jones type potential holds the haptic pen in place, allowing the user to feel competing repulsive and attractive forces as they transit the surface.   The spherical Coulomb potential, meanwhile, which mimics the force between two charged point particles, creates a `soft' sensation, with minimal pushback from the surface.  In general, the description of these forces is necessarily both subjective and qualitative in nature and all provided stable operation.

\section{Manipulation}

The controlled manipulation of atoms and molecules with an SPM is a powerful capability that enables the atom-by-atom engineering of nanosystems.\cite{bartels1997basic, stroscio2004controlling, crommie1993confinement, eigler1990positioning}  Haptic SPM control is especially useful for such a use case, as the operator can find the target atom/molecule, perform manipulation and interrogate the resulting structure without having to conduct a time-intensive raster scan after each manipulation attempt. To demonstrate this capability, we used haptic SPM control to manipulate the positions of CO molecules on a Cu(111) surface.  In these experiments, the system was operated in 'STM topography-based force' mode, as described in Sec. \ref{sec:stmread}.  The STM tip is first positioned on top of an individual CO molecule, and the lateral control of the haptic arm is momentarily paused to minimize uncontrolled movements.  Next, the STM tip is brought close to the molecule by increasing the setpoint current of the feedback and lowering the tunneling bias, with typical values of 10nA and 1mV, respectively.  Concurrently, the z-position of the tip is monitored for fluctuations, which indicate that the molecule is partially bound to the tip.\cite{stroscio2004controlling}  Lateral movement is then re-enabled, and the user can `drag' the molecule to the desired location upon which the tip is raised by increasing the bias and lowering the setpoint current to typical values of 50mV and 1nA, leaving the molecule on the surface.  

An example of this manipulation process is shown in Figure \ref{fig:manip}, where we plot $z^F_{pen}$ during the lateral sliding process of CO molecule. During the movement, the atomic lattice of the underlying fcc Cu(111) surface (with a lattice constant of 3.597 $\text{\AA}$ \cite{davey1925precision}) becomes resolved in STM measurements and can be perceived by the haptic pen due to the preferentially binding of the tops of the Cu atomic compared to the hollow site, which causes the CO to 'skip' between atomic sites.\cite{bartels1997basic, stroscio2004controlling, ternes2008force}  This process causes the CO molecule to dynamically alter the local conductivity during the sliding process, allowing the atomic lattice to be felt by the operator.  We found the process of manipulation to be reliable, and it was used to manipulate many molecules with high rates of success.  Images before and after the manipulation of 5 CO molecules are shown in Fig. \ref{fig:manip}c, with the latter spelling out `hi' in Unified English Braille.  We note that one distinct advantage of haptic-SPM is that during the manipulation process, it is easy to 'feel' around the surface to find molecules, without needing to re-scan the areas or to search for atoms by monitoring the height of the SPM tip visually.  This makes atomic manipulation less time consuming and more intuitive than conventional SPM operation.  

\section{Discussion and Outlook}

In general, the performance of our haptic-SPM was constrained by a number of factors, which we discuss as following.

One limitation of the specific haptic device used in this work is its relatively small range of motion in the $xy$ plane. Operations were limited to a 16 cm $\times$ 16 cm area, meaning that, for a large enough area of interest, small surface features were scaled down significantly and were more difficult to feel. A haptic device with a larger range of motion could address this issue by keeping the scale of horizontal movement constant, and increasing/decreasing the allowed range of motion for larger/smaller areas of interest, respectively. This would effectively allow the user to feel both large and small surface features at the same time, without sacrificing movement range for surface scale.

Another limitation is the discrepancy in the operating frequency of the Nanonis control software and the haptic pen. While the haptic pen updates at a rate of 1 kHz, the Nanonis control software is only capable of sending values to LabVIEW at a rate of 100 Hz. This results in the pen moving over many points before refreshing, limiting the pen from fully taking advantage of its 0.02 mm positional resolution. A faster control software operating frequency would yield significantly smoother operation, allowing larger areas and features to be felt at higher resolutions and speed. Ultimately, the control software's refresh rate is limited by the time constant of the feedback controller (typically $~30 \mu s$), meaning there is a maximal resolution capable of being achieved. 

Regardless of these technological limitations, our results show that the haptic controller described in this work provides a useful and intuitive way of interacting with a cryogenic, UHV STM/AFM. The variety of modes in which the device was successfully operated exemplifies the wide range of possible use-cases, and shows that the haptic SPM controller is an effective way to feel, simulate, and interact with nanoscale surfaces and features. 

Such a control system could be especially useful for educational and accessibility purposes. Tactile feedback can help develop more intuition about the operational principles of SPM technology than traditional sight-based methods, primarily due to the high correlation between tip-surface and pen-hand interactions. Haptic operations also provide near-instantaneous feedback when compared to obtaining time-intensive raster scans, facilitating  fine-grained operations like molecular manipulation. Scanning probe microscopy is essential for the study and engineering of nanoscale systems, and further refinement of haptic-SPM controllers will only serve to make interacting with such instruments more ergonomic and more accessible to the layperson.

\begin{acknowledgments}

Work by M. F. and R. A. was supported by the National
Science Foundation (NSF), Division of Materials Research
(RAISE DMR) under Award No. 1839199.  Work by W. A. B. and V. W. B. was supported by the Office of Naval Research under
Award No. N00014-20-1-235.

\end{acknowledgments}

\section*{Author Declarations}

\subsection*{Conflict of Interest}

The authors have no conflicts to disclose.

\subsection*{Author Contributions}

\textbf{M. Freeman and R. Applestone:} data curation, investigation, methodology, software, visualization, writing - original draft, writing - review and editing. 
\textbf{W. Behn:} data curation, investigation, supervision. 
\textbf{V. Brar:} conceptualization, funding acquisition, project administration, supervision, writing - original draft, writing - review and editing.

\section{Data Availability}

The data that support the findings of this study are available from the corresponding author upon reasonable request or from:\newline \url{https://github.com/HapticSPM/data}.

\bibliography{main}

\end{document}